\documentclass[journal]{IEEEtran}
\usepackage{amsmath, amssymb, latexsym, amsfonts, epsfig, stmaryrd}

\newcommand{\comment}[1]{}

\def\Rmac{{\cal R}_{MAC}}

\newtheorem{theorem}{Theorem}
\newtheorem{lemma}[theorem]{Lemma}

\newtheorem{definition}{Definition}

\renewcommand{\IEEEQED}{\IEEEQEDopen}

\begin{document}

\title{Variable Length Coding over the Two-User Multiple-Access Channel}

\author{\authorblockN{St\'ephane Musy\\
\authorblockA{School of Computer and Communication Sciences,
EPFL\\ CH-1015 Lausanne, Switzerland\\
%\authorblockA{EPFL --- I\&C --- LTHI\\ 
%CH-1015 Lausanne, Switzerland\\
%Email: \{stephane.musy,emre.telatar\}@epfl.ch\\[-3ex]}}
Email: stephane.musy@a3.epfl.ch}}
%\and
%\authorblockN{Emre Telatar}
%\authorblockA{School of Computer and Communication Sciences\\
%EPFL\\ CH-1015 Lausanne, Switzerland\\
%Email: emre.telatar@epfl.ch}
}
%\authorblockN{Leif Hanlen}
%\authorblockA{Wireless Signal Processing Program\\
%National ICT Australia\\
%Canberra ACT 0200, Australia\\
%Email: Leif.Hanlen@nicta.com.au} \and
%\authorblockN{Alex Grant}
%\authorblockA{Inst. for Telecommunications Research\\
%University of South Australia\\
%Mawson Lakes SA 5095, Australia\\
%Email: Alex.Grant@unsa.edu.au}
%}

% avoiding spaces at the end of the author lines is not a problem with
% conference papers because we don't use \thanks or \IEEEmembership
%Homer Simpson\authorrefmark{2},
%Montgomery Scott\authorrefmark{3} and
%Eldon Tyrell\authorrefmark{4}}
%Telephone: (800) 555--1212, Fax: (888) 555--1212}
%\authorblockA{\authorrefmark{4}Tyrell Inc., 123 Replicant Street, Los Angeles, California 90210--4321}}

\maketitle

{\def\thefootnote{}\footnotetext{The work presented in this paper 
was partially supported by the
National Competence Center in Research on Mobile Information and
Communication Systems (NCCR-MICS), a center supported by the Swiss
National Science Foundation under grant number 5005-67322.}}
\begin{abstract}
%This paper investigates the rates achievable by 
%variable length coding over a two-user multiple-access channel.
%We define the transmission rates from the
%perspective of the receiver and study the set of
%achievable rates when the receiver is allowed to decode
%each transmitted message at a different instant of time.
%We give an outer bound to this region,
%as well as examples of code that achieve this bound
%in particular settings.
For discrete memoryless multiple-access channels, 
we propose a general definition of variable length codes with
a measure of the transmission rates at the receiver side. 
This gives a receiver perspective on the multiple-access channel coding 
problem and allows us to characterize the region of achievable 
rates when the receiver is able to decode each transmitted message 
at a different instant of time. 
We show an outer bound on this region 
and derive a simple coding scheme that can achieve, in particular settings,
all rates within the region delimited by the outer bound. 
In addition, we propose a random variable length coding scheme 
that achieve the direct part of the block code capacity region 
of a multiple-access channel without requiring any agreement 
between the transmitters. 
\end{abstract}


\section{Introduction}
In this paper, we investigate the rates achievable
by using variable length codes over a two-user
multiple-access channel.
We let the codewords of each transmitter to be infinite 
sequences of input symbols%
\footnote{There are no feedback links, however,
in an implementation one can imagine a weak feedback
indicating when the receiver has made a decision.} 
and let the receiver decode each transmitted message
at some desired instant of time.%
\footnote{Note that both transmitters start to send their codeword at
the same instant of time.} 
The transmission ``rate'' of each message
is then defined from the perspective of the receiver,
as the information symbols transmitted per 
channel observation at the receiver.
Notice that in the usual sense these codes are rateless (or zero-rate),
here the transmission ``rate'' captures the
trade-off between the amount of information received with
the ``timeliness'' of the information.
This setting can be seen as a ``one-shot'' view on the
multiple-access communication problem as opposed
to a ``multi-shot'' view, where each transmitter has an
indefinite amount of information to simultaneously send to
the receiver, which is the view traditionally considered in 
network information theory. 
This approach may be useful to analyze scenarios where synchronous
users have {\em infrequent} messages to transmit.

Note that a definition of rates from the perspective 
of the receivers is made in \cite{Shul03} and \cite{SF00} 
to analyze broadcast channels 
where a common message is transmitted to several receivers.
Therein, the rate for each receiver is normalized by the time the receiver
needs to be ``online'' to reliably decode the message.
In this context, it is known that if the capacity 
achieving distribution is the same for each individual link, 
the maximum achievable transmission rate over each link 
can be simultaneously achieved.
A result that one can not reach with the classical
definitions of rates and block codes.
An effective way of achieving this when the receivers
are served by erasure channels is to use fountain codes, 
such as LT codes \cite{L02} or raptor codes \cite{S06}.
Notice that, an information theoretic treatment of 
fountain codes with a careful definition of rate
is done in \cite{STV07}.

In our setting, the following argument shows that, 
if we require that the receiver decode the transmitted messages
at the same instant of time, the set of achievable rates 
is the same for variable and fixed length codes.
To the contrary assume that such a code exists,
let $E[N]$ be its expected length, then 
by the law of large numbers the total length of $n$
successive transmissions is very likely to be less than
$n {\big (} E[N]+\epsilon {\big )}$. Thus, a fixed length
code of this length will achieve almost the same rate with
a small probability of error.%
\footnote{This argument can be formulated for any multiple-user channel.}
Therefore, the interesting problem is to characterize
the region of achievable rates when the receiver is 
allowed to decode the messages at different instants of time.

Here, we introduce a region of achievable rates that captures
the variability in the receiver decoding times and show
an outer bound on it.
This outer bound can be related to the block code capacity
region and quantify the possible gain over block codes
in terms of achievable rates.
Then, we present two examples of variable length codes
obtained by combination of block codes that achieve 
any rate pair within the region delimited by the outer bound,
in specific settings which are explicated later.
This argues that the gain in the achievable rates using
variable length codes comes only from the possibility for 
the receiver to decode the transmitted messages in 
non-overlapping periods of time.%
\footnote{Notice that the corresponding analysis for variable
length coding over a degraded broadcast channel in which independent
messages have to be transmitted to each receiver is done in \cite{M07}.}

To conclude, using random coding, we show the existence 
of a variable length code that achieves all rate pairs
within the direct part (without time-sharing) of the block 
code capacity region of a multiple-access channel, 
without requiring a previous agreement between the transmitter.%
\footnote{This means that no explicit or implicit agreement is made
between the transmitters, that is each transmitter acts as if
it were alone completly ignoring the other one.}
A result that one can not obtain using only block codes and
which might be interesting in a decentralized setting.

The next section provides the definition of a variable
length code for a multiple-access channel, along with
an associated region of achievable rates addressing the
possibility for the receiver to decode different instants
of time. 
In Section II, we show an outer bound on this region.
Then, in Section III, we relate the outer region 
formed by the outer bound to the
block code capacity region of a multiple-access channel,
and, in Section IV, we presents two examples of coding schemes 
based on block codes that achieve the outer region
in particular settings. 
Finally, in Section V, we explore
the set of rates achievable using variable length codes
with a random codebook and 
derive a decoding rule that achieves all rate pairs
within the direct part of the block code capacity region,  
without requiring any agreement between the transmitter.

\section{Definitions}
\begin{figure}[h]
\centering
\includegraphics[width=\linewidth]{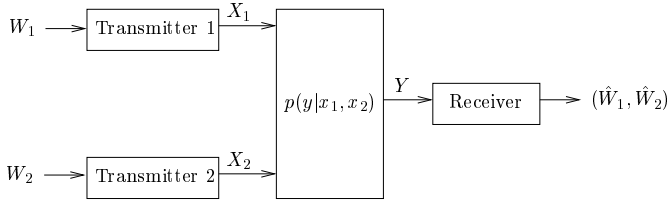}
 \caption{\small  Multiple-Access Channel.}\label{fig:MACblock}
\end{figure}

We consider a discrete memoryless multiple-access channel in which
two transmitters send independent information to a common receiver.
The channel model is illustrated in Figure \ref{fig:MACblock}. 
There are two sources, one producing a message $W_1 \in \{1,2,\dots,M_1\}$
and the other producing a message $W_2 \in \{1,2,\dots,M_2\}$.
The channel consists of two input alphabets ${\mathcal X_1}$ 
and ${\mathcal X_2}$, one output
alphabet ${\mathcal Y}$, and a probability transition 
function $p(y|x_1,x_2)$. By the memorylessness of the channel
we have, for any $n$, $p(y^n|x_1^n,x_2^n)=\Pi_{i=1}^{n} p(y_i|x_{1i},x_{2i})$,
where $x_1^n \in {\mathcal X_1}^n$, $x_2^n \in {\mathcal X_2}^n$ 
and $y^n \in {\mathcal Y}^n$.

\begin{figure}[h]
\centering
\includegraphics[totalheight=4.5cm]{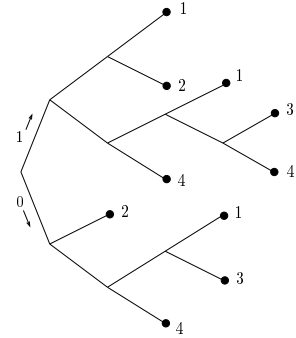}
 \caption{\small  Example of a tree associated with $g_1$
for a binary-output multiple-access channel with $M_1=4$.
The set of all received sequences for which a decision is made 
is represented by the leaves of a complete binary tree.
The decoder climbs the tree by going up or down whether it receives a one or
a zero, until it reaches a leaf and makes a decision accordingly.}
\label{fig:CodeTree}
\end{figure}

Let $N_1$ and $N_2$ be stopping times with respect to 
$\{Y_i\}_{i\geq 1}$, the sequence of received letters.
We define a $(M_1,M_2,N_1,N_2)$ 
{\em variable length code} as two sequences of mappings (encoders) 
$\{x_{1i}(W_1)\}_{i\geq 1}$ and $\{x_{2i}(W_2)\}_{i\geq 1}$, and two decoding 
functions (decoders) with respect to the decoding times $N_1$ and $N_2$,
\[g_1:{\mathcal Y}^{N_1} \rightarrow \{1,2,\dots,M_1\}\]
\noindent
and
\[g_2:{\mathcal Y}^{N_2} \rightarrow \{1,2,\dots,M_2\}.\] 
Note that ${\mathcal Y}^{N_1}$ and ${\mathcal Y}^{N_2}$
take values in the set of all finite sequences of channel output.
For deterministic stopping rules, we can represent
the set of all output sequences for which a decision is made,
at each decoder ($g_1$ and $g_2$), 
as the leaves of a complete $|{\mathcal Y}|$-ary tree.%
\footnote{A tree is said to be a complete $|{\mathcal Y}|$-ary 
tree if any vertex is either a leaf or has $|{\mathcal Y}|$
immediate descendants.}
The leaves have
a label from the set of messages. Each decoder starts
climbing the tree from the root. At each time
it chooses the branch that corresponds to the
received symbol. When a leaf is reached, the 
decoder makes a decision as indicated by the
label of the leaf (see Fig. \ref{fig:CodeTree} for an example).

Now, assuming that $(W_1,W_2)$ are uniformly distributed over 
$\{1,2,\dots,M_1\} \times \{1,2,\dots,M_2\}$,
we let the average probability of error to be the probability that
the decoded message pair is not equal to the transmitted one, i.e.,
\[ P_{e}=Pr\{g_1(Y^{N_1})\neq W_1 \mbox{ or } g_2(Y^{N_2})\neq W_2\ \},\]
and we define the transmission rates from the perspective of 
the receivers as $\frac{\log M_1}{E[N_1]}$ 
and $\frac{\log M_2}{E[N_2]}$.%
\footnote{Note that the expectation $E[N_1]$ and $E[N_2]$ 
are taken over the channel realizations and 
over the pair of messages $(W_1,W_2)$.}
Notice that this definition of rate is usually made
for variable length coding over a single-user channel,
see, e.g., \cite{Burn76,Tcham05}.
However, this is a particular choice that measures the rate
by the amount of information received over the average transmission time,
one can imagine other definitions that may lead to different results.

\begin{definition}
\label{def:achiv}
A rate pair $(R_1,R_2)$ is said to be {\em achievable} 
for the multiple-access channel if for all $\epsilon >0$,
there exists a $(M_1,M_2,N_1,N_2)$ variable length code with
$\frac{\log M_1}{E[N_1]}\geq R_1$, $\frac{\log M_2}{E[N_2]}\geq R_2$
and $P_{e}<\epsilon$.
\end{definition}

The {\em capacity region} of the multiple-access channel is the closure
of the set of achievable rates.%
\footnote{Here we consider the average probability of error.
To use $\hat P_e=\max_{w_1,w_2} \text{Pr}\{g_1(Y^N_1)=w_1 \text{ or }
g_2(Y^{N_2})=w_2|W_1=w_1,W_2=w_2\}$ would in general lead
to a different capacity region, as noticed in \cite{D78}.}
Observe that with this definition the capacity region is simply given
by the rectangle $[0,C_1]\times[0,C_2]$,
where $C_1 \triangleq \max_{p(x_1)p(x_2)} I(X_1;Y|X_2)$ and
$C_2 \triangleq \max_{p(x_1)p(x_2)} I(X_2;Y|X_1)$ are
the supremum of all achievable rates in each individual link.
As previously observed in \cite{C98}, any rate pair in this region 
can be achieved by sending the messages of each user
in a separated period of time, and by making the ratio $E[N_1]/E[N_2]$
approach zero (or infinity). %Therefore, we are interested by a notion of capacity 
Thereby requiring that one user have infinitely more information
to transmit than the other.

As mentioned in the introduction, here we want to consider scenarios
where each user has infrequent messages to transmit.
%For example, we can consider a synchronous network
%with infrequently occurring packets.
Thus, we are more interested to characterize  
the region of achievable rates for bounded values of the ratio
$E[N_1]/E[N_2]$ and capture the variability on the
receiver decoding times, this leads us to
consider the following region:

\begin{definition}
Let $N\triangleq\min(N_1,N_2)$,
we denote by ${\cal C}_{r_1,r_2}$ the set of rates
achievable by using variable length codes for which
$\frac{E[N]}{E[N_1]}\geq r_1$, $\frac{E[N]}{E[N_2]} \geq r_2 \triangleq sr_1$, 
with $0 \leq r_1,r_2 \leq 1$.
\end{definition}

This definition precludes the possibility that the receiver decodes
one transmitted message in a short period of time while the other one
takes a large period of time,
the ratio between the two average decoding times being governed
by the values of $r_1$ and $r_2$. 
The justification for the particular formulation of the restrictions
imposed on $E[N_1]$ and $E[N_2]$
comes from the outer bound that we found on this region,
this bound is presented in the next section.
Section V will then describe coding schemes 
based on block codes that achieve the
outer region when additional constraints are imposed on $r_1$ and $r_2$. 

\section{Outer Region} 
In order to prove our outer bound on 
${\cal C}_{r_1,r_2}$ we need two lemmas,
which gives lower bounds on the mutual
information of interest in terms of single letter
expressions.

\begin{lemma}
\label{maclemma2}
The following inequalities hold:
\begin{align*}
I(W_1;Y^N|W_2) &\leq E[N]I(X_1;Y|X_2,Q) + \log(eE[N])\\
I(W_2;Y^N|W_1) &\leq E[N]I(X_2;Y|X_1,Q) + \log(eE[N])\\
I(W_1,W_2;Y^N) &\leq E[N]I(X_1,X_2;Y|Q) + \log(eE[N]),
\end{align*}
for some joint distribution $p(q)p(x_1|q)p(x_2|q)p(y|x_1,x_2)$.
\end{lemma} 

\begin{IEEEproof}
Let $\lambda_i = 1\{N\geq i\}$,% 
\footnote{Where $1\{N\geq i\}$ is equal to 1 if $N\geq i$
and equal to 0 otherwise. Also, we define $A_i\lambda_i$
as being equal to $A_i$ if $N\geq i$ and equal 
to $\aleph$ otherwise, where $\aleph$ denotes a
symbol distinct from any of the letters in $({\mathcal X_1},{\mathcal X_2},
{\mathcal Y})$,
and $A_i$ can be either $X_{1i}$, $X_{2i}$ or $Y_i$.}
then, from the chain rule for mutual information, we have
\begin{align*}
I(W_1;Y^N|W_2) 
&= I(W_1;Y_1\lambda_1,\lambda_1,\cdots,Y_n\lambda_n,\lambda_n,\cdots|W_2)\\
&= I(W_1;\lambda_1|W_2) + I(W_1;Y_1\lambda_1|\lambda_1,W_2) + \cdots\\
&\quad + I(W_1;\lambda_n|(Y\lambda)^{n-1},\lambda^{n-1},W_2)\\
&\quad + I(W_1;Y_n\lambda_n|(Y\lambda)^{n-1},\lambda^n,W_2) + \cdots\\
&= \sum_{i=1}^{\infty} I(W_1;\lambda_i|(Y\lambda)^{i-1},\lambda^{i-1},W_2)\\
&\quad + \sum_{i=1}^{\infty} I(W_1;Y_i\lambda_i|(Y\lambda)^{i-1},\lambda^i,W_2).
\end{align*}
The first summation can be upper bounded as
\begin{align*}
\sum_{i=1}^{\infty} I(W_1;\lambda_i|(Y\lambda)^{i-1},\lambda^{i-1},W_2)
&\leq \sum_{i=1}^{\infty} H(\lambda_i|\lambda^{i-1})\\
&= H(\lambda_1,\lambda_2,\cdots)\\
&= H(N)\\
&\leq \log(eE[N]),
\end{align*}
where we use the fact that conditioning reduces entropy, and
the last inequality is proved in \cite{C73} 
and \cite[\textsection 1.3]{CK81},
for any non-negative discrete random variable,
using the log sum inequality.
\\
\\
For the second summation, we can write
\begin{align*}
&I(W_1;Y_i\lambda_i|(Y\lambda)^{i-1},\lambda^i,W_2)\\
&= H(Y_i\lambda_i|(Y\lambda)^{i-1},\lambda^i,W_2)
- H(Y_i\lambda_i|(Y\lambda)^{i-1},\lambda^i,W_2,W_1)\\
&\stackrel{(a)}{\leq} H(Y_i\lambda_i|X_{2i}\lambda_i,\lambda_i)
- H(Y_i\lambda_i|(Y\lambda)^{i-1},\lambda^i,W_2,W_1)\\
&\stackrel{(b)}{=} H(Y_i\lambda_i|X_{2i}\lambda_i,\lambda_i)
- H(Y_i\lambda_i|X_{1i}\lambda_i,X_{2i}\lambda_i,\lambda_i)\\
&= \text{Pr}(\lambda_i=1){\big[} H(Y_i|X_{2i},\lambda_i=1)
- H(Y_i|X_{1i},X_{2i},\lambda_i=1) {\big]}\\
&= \text{Pr}(N\geq i) I(X_{1i};Y_i|X_{2i},\lambda_i=1),
\end{align*}
where $(a)$ follows, since conditioning reduces entropy and
$X_{2i}$ is a function of $W_2$. In $(b)$ 
we remark that knowing $\lambda_i$, $Y_i\lambda_i$ is
independent of the past values $\{\lambda_j\}_{j<i}$, and 
that $(X_{1i},X_{2i})$ is a function of $(W_1,W_2)$
and then given $(X_{1i},X_{2i})$, $Y_i$ 
is independent of $(W_1,W_2)$ and of the past received values.
The other equalities follow by definition of the corresponding quantities.
\\
\\
Next, observe that $p(y_i|x_{1i},x_{2i},\lambda_i=1)=p(y_i|x_{1i},x_{2i})$,
thus $I(X_{1i};Y_i|X_{2i},\lambda_i=1)=I(X_{1i};Y_i|X_{2i})$,
with $p(x_{1i})\triangleq p(x_{1i}|\lambda_i=1)$ and
$p(x_{2i})\triangleq p(x_{2i}|\lambda_i=1)$. 
Hence, we get 
\begin{align*}
&\sum_{i=1}^{\infty} I(W_1;Y_i\lambda_i|(Y\lambda)^{i-1},\lambda^i,W_2)\\
&\leq \sum_{i=1}^{\infty}\text{Pr}(N\geq i)I(X_{1i};Y_i|X_{2i})\\
&= E[N]\sum_{i=1}^{\infty}\frac{\text{Pr}(N\geq i)}{E[N]}I(X_{1i};Y_i|X_{2i}).
\end{align*}
Now let $a_i=\frac{\text{Pr}(N\geq i)}{E[N]}$, 
note that $a_i\geq 0$ for all $i$, and $\sum_i a_i=1$.
Thus, we can define an integer random variable $Q$ by setting $\text{Pr}(Q=i)=a_i$,
for all $i\in\{1,2,\dots\}$.
Using this, the preceding equation becomes
\begin{align*}
&\sum_{i=1}^{\infty} I(W_1;Y_i\lambda_i|(Y\lambda)^{i-1},\lambda^i,W_2)\\
&= E[N]\sum_{i=1}^{\infty} \text{Pr}(Q=i) I(X_{1Q};Y_Q|X_{2Q},Q=i)\\
&= E[N]I(X_1;Y|X_2,Q),
\end{align*}
where $X_1\triangleq X_{1Q}$, $X_2\triangleq X_{2Q}$ 
and $Y\triangleq Y_Q$ are new random variables
whose distributions depend on $Q$ in the same way as
the distributions of $X_{1i}$, $X_{2i}$ and $Y_i$ depend on $i$.
Notice that $Q\rightarrow (X_1,X_2)\rightarrow Y$ forms 
a Markov chain. 
Therefore, we obtain
\begin{align*}
I(W_1;Y^N|W_2) &\leq E[N]I(X_1;Y|X_2,Q) + \log(eE[N]),
\end{align*}
for some joint distribution $p(q)p(x_1|q)p(x_2|q)p(y|x_1,x_2)$.
\\
\\
The second inequality follows in a symmetric way. For the last one,
we proceed in the same manner, consider
\begin{align*}
I(W_1,W_2;Y^N) 
&= I(W_1,W_2;Y_1\lambda_1,\lambda_1,\cdots,Y_n\lambda_n,\lambda_n,\cdots)\\
%&= I(W_1,W_2;\lambda_1) + I(W_1,W_2;Y_1\lambda_1|\lambda_1)\\
%&\quad + \cdots + I(W_1,W_2;\lambda_n|(Y\lambda)^{n-1},\lambda^{n-1})\\
%&\quad + I(W_1,W_2;Y_n\lambda_n|(Y\lambda)^{n-1},\lambda^n) + \cdots\\
&= \sum_{i=1}^{\infty} I(W_1,W_2;\lambda_i|(Y\lambda)^{i-1},\lambda^{i-1})\\
&\quad + \sum_{i=1}^{\infty} I(W_1,W_2;Y_i\lambda_i|(Y\lambda)^{i-1},\lambda^i).
\end{align*}
As before, the first summation can be upper bounded as
\begin{align*}
\sum_{i=1}^{\infty} I(W_1,W_2;\lambda_i|(Y\lambda)^{i-1},\lambda^{i-1})
%&\leq \sum_{i=1}^{\infty} H(\lambda_i|\lambda^{i-1})\\
%&= H(\lambda_1,\lambda_2,\cdots)\\
%&= H(N)\\
&\leq \log(eE[N]).
\end{align*}
%where the last inequality is proved in \cite{C73}
%and \cite[\textsection 1.3]{CK81},
%as mentioned before.
\noindent
For the second summation, we have
\begin{align*}
&I(W_1,W_2;Y_i\lambda_i|(Y\lambda)^{i-1},\lambda^i)\\
&= H(Y_i\lambda_i|(Y\lambda)^{i-1},\lambda^i)
- H(Y_i\lambda_i|(Y\lambda)^{i-1},\lambda^i,W_1,W_2)\\
&\leq H(Y_i\lambda_i|\lambda_i)
- H(Y_i\lambda_i|X_{1i}\lambda_i,X_{2i}\lambda_i,\lambda_i)\\
&= \text{Pr}(\lambda_i=1){\big[} H(Y_i|\lambda_i=1)
- H(Y_i|X_{1i},X_{2i},\lambda_i=1) {\big]}\\
&= \text{Pr}(N\geq i) I(X_{1i},X_{2i};Y_i|\lambda_i=1),
\end{align*}
since $(X_{1i},X_{2i})$ is a function of $(W_1,W_2)$, and
given $(X_{1i},X_{2i})$, $Y_i$ is independent of the past 
received values. % and of the event $\{N\geq i\}$.
\\
\\
Then, observe that 
$p(y_i|x_{1i},x_{2i},\lambda_i=1)=p(y_i|x_{1i},x_{2i})$,
thus $I(X_{1i},X_{2i};Y_i|\lambda_i=1)=I(X_{1i};Y_i|X_{2i})$,
with $p(x_{1i})\triangleq p(x_{1i}|\lambda_i=1)$ and
$p(x_{2i})\triangleq p(x_{2i}|\lambda_i=1)$. 
Hence, we get 
\begin{align*}
&\sum_{i=1}^{\infty} I(W_1,W_2;Y_i\lambda_i|(Y\lambda)^{i-1},\lambda^i)\\
&\leq \sum_{i=1}^{\infty}\text{Pr}(N\geq i)I(X_{1i},X_{2i};Y_i)\\
&= E[N]\sum_{i=1}^{\infty}\frac{\text{Pr}(N\geq i)}{E[N]}I(X_{1i},X_{2i};Y_i).
\end{align*}
Now, as done before, let $a_i=\frac{\text{Pr}(N\geq i)}{E[N]}$, 
and define an integer random variable $Q$ by setting $\text{Pr}(Q=i)=a_i$,
for all $i\in\{1,2,\dots\}$.
Using this, the preceding equation becomes
\begin{align*}
&\sum_{i=1}^{\infty} I(W_1,W_2;Y_i\lambda_i|(Y\lambda)^{i-1},\lambda^i)\\
&= E[N]\sum_{i=1}^{\infty} \text{Pr}(Q=i) I(X_{1Q},X_{2Q};Y_Q|Q=i)\\
&= E[N]I(X_1,X_2;Y|Q),
\end{align*}
where $X_1\triangleq X_{1Q}$, $X_2\triangleq X_{2Q}$ 
and $Y\triangleq Y_Q$ are random variables
whose distributions depend on $Q$ in the same way as
the distributions of $X_{1i}$, $X_{2i}$ and $Y_i$ depend on $i$.
Notice that $Q\rightarrow (X_1,X_2)\rightarrow Y$ forms 
a Markov chain.
\\
\\ 
Therefore, we obtain
\begin{align*}
I(W_1,W_2;Y^N) &\leq E[N]I(X_1,X_2;Y|Q) + \log(eE[N]),
\end{align*}
for some joint distribution $p(q)p(x_1|q)p(x_2|q)p(y|x_1,x_2)$.
%\qed
\end{IEEEproof}

We show the proof of the next lemma in appendix, the
main ideas being presented in the previous lemma. 
\begin{lemma}
\label{maclemma3}
We have the following inequalities:
\begin{align*}
I(W_1;Y_{N+1}^{N_1}|Y^N,W_2)&\leq E[N_1-N] C_1 + \log(eE[N_1-N])\\
I(W_2;Y_{N+1}^{N_2}|Y^N,W_1)&\leq E[N_2-N] C_2 + \log(eE[N_2-N]).
\end{align*}
\end{lemma} 

\begin{IEEEproof}
See Appendix \ref{app:lem}.
\end{IEEEproof}

Notice that in these lower bounds the additional terms
corresponding to the information provided by the length
of the codewords are sublinear in the average decoding times.
This is an interesting fact that we use to show our outer bound 
on the region of achievable rates ${\cal C}_{r_1,r_2}$,
given by the following theorem. 
\begin{theorem}
%\emph{(Feedback capacity)}
\label{th:convmac}
\emph{(Outer bound)} 
%$\mathcal{C}_r = \{R_1\leq I(X;Y|U),R_2\leq rI(U;Z)+(1-r)C_2\}
%$\cup \{R_1\leq r I(X;Y|u)+(1-r)C_1,R_2\leq I(U;Z)\}$
%Let us denote by ${\cal C}_{r_1,r_2}$ the set of rates
%achievable by using variable length codes for which
%$\frac{E[N]}{E[N_1]}\geq r_1$, $\frac{E[N]}{E[N_2]} \geq r_2 \triangleq sr_1$, 
%with $0 < r_1,r_2 \leq 1$. 
%and $r_2\leq s \leq \frac{1}{r_1}$.
%Then, 
Any rate pair $(R_1,R_2)\in{\cal C}_{r_1,r_2}$
must satisfy
%Let $r_1=\frac{E[N]}{E[N_1]}$, $r_2=\frac{E[N]}{E[N_2]}$
%and  $s=\frac{r_2}{r_1}=\frac{E[N_1]}{E[N_2]}$. 
%For a variable length code with stopping time $N_1$ and $N_2$,
%all achievable rate pairs $(R_1,R_2)$ must satisfy
%
%\[
%E[N_2]R_2 + \lambda E[N_1]R_1\leq E[\min(N_1,N_2)]{\big[}C(\lambda)-C_2 
%- \lambda C_1{\big]}+ E[N_2] C_2 +\lambda E[N_1] C_1,
%\]
%for all $\lambda>0$.
%\begin{align*}
%R_1&\leq \frac{1}{E[N_1]}E[\min(N_1,N_2)I_{\min(N_1,N_2)}(X_1;Y|X_2,Q)] + (1-r_1)C_1\\
%R_2&\leq \frac{1}{E[N_2]}E[\min(N_1,N_2)I_{\min(N_1,N_2)}(X_2;Y|X_1,Q)] + (1-r_2)C_2\\
%s R_1 + R_2 &\leq \frac{1}{E[N_2]}E[\min(N_1,N_2)I_{\min(N_1,N_2)}(X_1,X_2;Y|Q)] 
%+ s(1-r_1)C_1 + (1-r_2)C_2 ,
%\end{align*} 
\begin{align*}
R_1&\leq r_1 I(X_1;Y|X_2,Q) + (1-r_1)C_1\\
R_2&\leq r_2 I(X_2;Y|X_1,Q) + (1-r_2)C_2\\
s R_1 + R_2 &\leq r_2 I(X_1,X_2;Y|Q) + s(1-r_1)C_1 + (1-r_2)C_2, 
\end{align*} 
for some joint distribution $p(q)p(x_1|q)p(x_2|q)p(y|x_1,x_2)$,
with $|{\cal Q}|\leq 2$.
\end{theorem}

\begin{IEEEproof}
Let $W_i$ be uniformly distributed over $\{1,2,\dots,M_i \}$, $i=1,2$.
Then,
\begin{align*}
&I(W_1,W_2;Y^{\max(N_1,N_2)})\\
&=H(W_1,W_2) - H(W_1,W_2|Y^{\max(N_1,N_2)})\\
&=E[N_1]R_1 + E[N_2]R_2 - H(W_1,W_2|Y^{\max(N_1,N_2)}),
\end{align*}
and
\begin{align*}
I(W_1;Y^{N_1}|W_2)&=H(W_1|W_2) - H(W_1|Y^{N_1},W_2)\\
&\geq E[N_1]R_1 - H(W_1|Y^{N_1}),
\end{align*}
and
\begin{align*}
I(W_2;Y^{N_2}|W_1)&=H(W_2|W_1) - H(W_2|Y^{N_2},W_1)\\
&\geq E[N_2]R_2 - H(W_2|Y^{N_2}).
\end{align*}

\noindent
From Fano's inequality, we have
\begin{align*}
E[N_1](R_1-\epsilon) &\leq I(W_1;Y^{N_1}|W_2)\\
E[N_2](R_2-\epsilon) &\leq I(W_2;Y^{N_2}|W_1)\\
E[N_1](R_1-\epsilon) + E[N_2](R_2-\epsilon) &\leq I(W_1,W_2;Y^{\max(N_1,N_2)}), 
\end{align*}
where $\epsilon \rightarrow 0$ as $P_e \rightarrow 0$.
\\
\\
Applying the chain rule for mutual information
and remembering that $N=\min(N_1,N_2)$, 
we can write
\begin{align*}
I(W_1;Y^{N_1}|W_2) %&\leq I(W_1;Y^{N_1},N|W_2)\\
%&= I(W_1;N|W_2) + I(W_1;Y^{N_1}|N,W_2)\\
%&= I(W_1;N|W_2)+I(W_1;Y^N|N,W_2)+I(W_1;Y_{N+1}^{N_1}|N,Y^{N},W_2)\\
&= I(W_1;Y^{N}|W_2)+ I(W_1;Y_{N+1}^{N_1}|Y^{N},W_2),
\end{align*}
with the convention that $Y^{N}_{N+1}=\emptyset$.
\\
\\
Then, using Lemma \ref{maclemma2} and Lemma \ref{maclemma3}, we get
\begin{align*}
I(W_1;Y^{N_1}|W_2) &\leq E[N]I(X_1;Y|X_2,Q) + E[N_1-N]C_1\\
&\quad + \log(eE[N]) + \log(eE[N_1-N]),
\end{align*}
for some joint distribution $p(q)p(x_1|q)p(x_2|q)p(y|x_1,x_2)$.
\\
\\
In a symmetric way, we obtain
\begin{align*}
I(W_2;Y^{N_2}|W_1) &\leq E[N]I(X_2;Y|X_1,Q) + E[N_2-N]C_2\\
&\quad + \log(eE[N]) + \log(eE[N_2-N]),
\end{align*}
for some joint distribution $p(q)p(x_1|q)p(x_2|q)p(y|x_1,x_2)$.
\\
\\
Now, using the chain rule for mutual information, we have
\begin{align*}
&I(W_1,W_2;Y^{\max(N_1,N_2)})\\  %&\leq I(W_1,W_2;Y^{\max(N_1,N_2)},N)\\
%&= I(W_1,W_2;N)+I(W_1,W_2;Y^{\max(N_1,N_2)}|N)\\
%&= I(W_1,W_2;N) + I(W_1,W_2;Y^N|N) 
%+ I(W_1,W_2;Y_{N+1}^{\max(N_1,N_2)}|N,Y^N)\\
&= I(W_1,W_2;Y^N) + I(W_1,W_2;Y_{N+1}^{\max(N_1,N_2)}|Y^N)\\
&= I(W_1,W_2;Y^N) + I(W_1,W_2;Y_{N+1}^{\max(N_1,N_2)}|N,Y^N).
\end{align*}
Lemma \ref{maclemma2} implies
\begin{align*}
I(W_1,W_2;Y^N)&\leq E[N]I(X_1,X_2;Y|Q) + \log(eE[N]),
\end{align*} 
for some joint distribution $p(q)p(x_1|q)p(x_2|q)p(y|x_1,x_2)$.
For the second term, the following holds
\begin{align*}
&I(W_1,W_2;Y_{N+1}^{\max(N_1,N_2)}|N,Y^N)\\
&=\text{Pr}(N=N_1) I(W_1,W_2;Y_{N_1+1}^{N_2}|N=N_1,Y^{N_1})\\
& \quad + \text{Pr}(N=N_2) I(W_1,W_2;Y_{N_2+1}^{N_1}|N=N_2,Y^{N_2}) \\
&=\text{Pr}(N_1\leq N_2) I(W_1,W_2;Y_{N_1+1}^{N_2}|N=N_1,Y^{N_1})\\ 
&\quad + \text{Pr}(N_2<N_1) I(W_1,W_2;Y_{N_2+1}^{N_1}|N=N_2,Y^{N_2}),
%&=I(W_1,W_2;Y_{N+1}^{N_1}|N,Y^N) + I(W_1,W_2;Y_{N+1}^{N_2}|N,Y^N),
\end{align*}
with
\begin{align*}
I(W_1,W_2&;Y_{N_1+1}^{N_2}|N=N_1,Y^{N_1})\\ 
&= I(W_2;Y_{N_1+1}^{N_2}|N=N_1,Y^{N_1},W_1)\\
&\quad +I(W_1;Y_{N_1+1}^{N_2}|N=N_1,Y^{N_1})\\
&= I(W_2;Y_{N_1+1}^{N_2}|N=N_1,Y^{N_1},W_1)\\
&\quad + H(W_1|N=N_1,Y^{N_1})\\
&\quad - H(W_1|N=N_1,Y^{N_2}).
\end{align*}
Since at time $N_1$ the receiver decodes $W_1$, 
we can apply Fano's inequality, yielding
\begin{align*}
I(W_1,W_2&;Y_{N_1+1}^{N_2}|N=N_1,Y^{N_1})\\
&\leq I(W_2;Y_{N_1+1}^{N_2}|N=N_1,Y^{N_1},W_1)+E[N_1]\epsilon, 
\end{align*}
where $\epsilon\rightarrow 0$ as $P_e \rightarrow 0$.
By symmetry, we have
\begin{align*}
I(W_1,W_2&;Y_{N_2+1}^{N_1}|N=N_2,Y^{N_2})\\
&\leq I(W_1;Y_{N_2+1}^{N_1}|N=N_2,Y^{N_2},W_2)+E[N_2]\epsilon. 
\end{align*}
Hence,
\begin{align*}
I(W_1,W_2&;Y_{N+1}^{\max(N_1,N_2)}|N,Y^N)\\
&\leq\text{Pr}(N_1\leq N_2)I(W_2;Y_{N_1+1}^{N_2}|N=N_1,Y^{N_1},W_1)\\
&\quad + \text{Pr}(N_2<N_1)I(W_1;Y_{N_2+1}^{N_1}|N=N_2,Y^{N_2},W_2)\\
&\quad + E[N_1]\epsilon + E[N_2]\epsilon\\
&= I(W_2;Y_{N+1}^{N_2}|N,Y^{N},W_1)\\
&\quad + I(W_1;Y_{N+1}^{N_1}|N,Y^{N},W_2)\\
&\quad + E[N_1]\epsilon + E[N_2]\epsilon\\
&\leq E[N_2-N]C_2 + E[N_1-N]C_1\\
&\quad + \log(eE[N_2-N]) + \log(eE[N_1-N])\\
&\quad + E[N_1]\epsilon + E[N_2]\epsilon,
\end{align*}
where we use Lemma \ref{maclemma3} to obtain the last inequality.
\\
\\
Putting things together, we get
\begin{align*}
E[N_1](R_1-\epsilon)&\leq E[N] I(X_1;Y|X_2,Q) + E[N_1-N]C_1\\
&\quad + \log(eE[N]) + \log(eE[N_1-N])
\end{align*}

\begin{align*}
E[N_2](R_2-\epsilon)&\leq E[N] I(X_2;Y|X_1,Q) + E[N_2-N]C_2\\
&\quad + \log(eE[N]) + \log(eE[N_2-N])
\end{align*}

\begin{align*}
E[N_1](R_1-\epsilon) &+ E[N_2](R_2-\epsilon)\\
&\leq E[N] I(X_1,X_2;Y|Q)+ E[N_1-N]C_1\\ 
&\quad + E[N_2-N]C_2+\log(eE[N])\\
&\quad +\log(eE[N_1-N])+\log(eE[N_2-N])\\
&\quad +E[N_1]\epsilon+E[N_2]\epsilon, 
\end{align*} 
for some joint distribution $p(q)p(x_1|q)p(x_2|q)p(y|x_1,x_2)$.
\\
\\ 
Dividing by $E[N_1]$
in the first inequality
and by $E[N_2]$ in the second and in the last inequality,
then letting  $E[N_1]\rightarrow \infty$ and  
$E[N_2]\rightarrow \infty$ with $\frac{E[N_1]}{E[N_2]}=s$  
gives the statement of the theorem.
The upper bound on the cardinality of ${\cal Q}$ follows
from convex analysis. 

%\hfill
%\qed
\end{IEEEproof}

In the previous proof we let the expected
decoding times be arbitrary large, but in regards of
our definition of achievability (Definition \ref{def:achiv})
it is not sure that this is needed in order to achieve 
an arbitrary low probability of error.%
\footnote{Here the concatenation argument traditionally made with block codes
does not work.}
For channels with a zero-error capacity equal to zero,
Appendix \ref{app:achiv} gives an heuristic argument
showing that this is indeed required.
However, observe that variable length codes can increase
the zero-error capacity of a channel (for example, one can consider the binary
erasure channel), thus this is not just a technicality.
%and needs to be further investigated.

\section{Comments on the Outer Region}
Let ${\cal R}_{MAC}$ denote the block code capacity 
region of a multiple-access channel, which can be stated
as the union of all pairs $(R_1,R_2)$ satisfying%
\footnote{For a careful definition and analysis of block
codes and multiple-access channels, the reader is
referred to \cite{TC06} and the references therein.}
\begin{align*}
R_1 &\leq I(X_1;Y|X_2,Q)\\
R_2 &\leq I(X_2;Y|X_1,Q)\\
R_1 + R_2 &\leq I(X_1,X_2;Y|Q),
\end{align*}
for some joint distribution $p(q)p(x_1|q)p(x_2|q)p(y|x_1,x_2)$,
with $|{\cal Q}|\leq 2$.

For a given $r_1$ and $r_2$,
let us rewrite the region defined by the outer bound of
the previous theorem as 
the union of all $(R^{\prime}_1,R^{\prime}_2)$ pairs satisfying
\begin{align*}
R^{\prime}_1&\leq r_2 I(X_1;Y|X_2,Q) + s(1-r_1)C_1\\
R^{\prime}_2&\leq r_2 I(X_2;Y|X_1,Q) + (1-r_2)C_2\\
R^{\prime}_1 + R^{\prime}_2 &\leq r_2 I(X_1,X_2;Y|Q) + s(1-r_1)C_1 + (1-r_2)C_2,
\end{align*}
for some joint distribution $p(q)p(x_1|q)p(x_2|q)p(y|x_1,x_2)$,
with $|{\cal Q}|\leq 2$. We have just set $R^{\prime}_1=s R_1$ 
and $R^{\prime}_2=R_2$ in the region of the theorem.
Denote it by ${\cal R}$.
From these expressions, we have immediately
that $(R^{\prime}_1,R^{\prime}_2)\in {\cal R}$ is
equivalent to
\begin{align*}
\frac{1}{r_2}{\big(}R^{\prime}_1-s(1-r_1)C_1,R^{\prime}_2-(1-r_2)C_2{\big)}
\in \Rmac.
\end{align*}

\begin{figure}[h]
\centering
\includegraphics{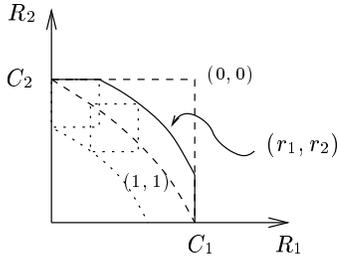}
\caption{\small  Example of an outer region  with an
arbitrary $(r_1,r_2)$.
The dashed line with $r_1=1$ and $r_2=1$ represents
the block code capacity region of a multiple-access channel.
The dotted lines show the construction of the outer region.}
\label{fig:bcregion22bw}
\end{figure}

Therefore, the region given by Theorem \ref{th:convmac} can be seen as
a contraction by $(r_1,r_2)$ of the block code capacity region 
of a multiple-access channel followed 
by an extension of $((1-r_1)C_1,(1-r_2)C_2)$.
This is illustrated in Fig. \ref{fig:bcregion22bw}.
One can also remark that, when $(r_1,r_2)=(1,1)$ the outer region
is equal to the block code capacity region $\Rmac$, and
for $(r_1,r_2)=(0,0)$ we recover the full rectangle
$[0,C_1]\times[0,C_2]$. 

Finally, let us emphasize that ${\cal C}_{r_1,r_2}$ is defined
for variable length codes with a certain $r_1$ and $r_2$, 
and that no bounds on the possible values of these ratios are given here. 
However the existence of a coding scheme with any desired 
$r_1$ and $r_2$ is not guaranteed.
In the next section we specify the outer region 
when some restriction on $E[N_1]$, $E[N_2]$ and $E[N]$
are imposed and show explicit coding schemes that
achieve the outer region in these particular cases.

\section{Achievability and Coding Schemes}
Let us first restrict the analysis to coding schemes for which
the receiver never (or with a negligible probability) 
decodes the message from the first transmitter
after the message coming from the second transmitter, that is $E[N]=E[N_1]$
or equivalently $r_1=1$.  
In this case, the outer bound of Theorem \ref{th:convmac} can be written
as, any rate pair $(R_1,R_2)\in{\cal C}_{1,r_2}$ must satisfy
\begin{align*}
R_1&\leq I(X_1;Y|X_2,Q)\\
R_2&\leq s I(X_2;Y|X_1,Q) + (1-r_2)C_2\\
r_2R_1 + R_2 &\leq r_2 I(X_1,X_2;Y|Q) + (1-r_2)C_2,
\end{align*}
for some joint distribution $p(q)p(x_1|q)p(x_2|q)p(y|x_1,x_2)$
with $|{\cal Q}|\leq 2$,% 
\footnote{Henceforth we will omit to mention the cardinality
bound on ${\cal Q}$.}
and where $0 \leq r_2 \leq 1$.

\begin{figure}[h]
%\vspace{.5\baselineskip}
\centering
\includegraphics{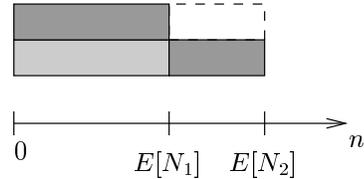}
\caption{Example of codewords formed by the concatenation
of two block codes.
The top (resp. bottom) line illustrates the codeword of 
the first (resp. second) transmitter. 
The filled intensity of a block representing a codeword
is proportional to the information rate of 
the corresponding code.}\label{fig:bcode}
\end{figure}

As the following construction will show,
any rate pair in the region delimited by this outer bound 
can be achieved by using a (sequence of) 
concatenation of two (multiple-access) block codes.
For some $\epsilon>0$, generate one block code of length $E[N_1]$ 
and rates $(R_1^*,R_2^*)\in{\cal R}_{MAC}$,
and one of length $E[N_2]-E[N_1]$ and rates $(0,C_2-\epsilon)$, that is
the first transmitter send the input symbol that allows
the second transmitter to send at its maximum rate 
(see Fig. \ref{fig:bcode}).

Denote by ${\cal V}^1$ the $(M_1,M_2,N^1_1,N^1_2)$ variable length code 
obtained by the concatenation of these two block codes,
this means that we let the codewords be formed by the Cartesian product
of the respective codebooks,%
\footnote{To be rigorous
we should add to each codeword an infinite sequence 
of arbitrary input symbols.}
and that the decoding functions are equal to the corresponding
block code decoding functions with respect to the
fixed stopping times $N_1^1$ and $N_2^1$,
which are given by
\begin{align*}
N^1_1&=E[N_1]=\frac{\log M_1}{R_1^*}\text{  a.s.}\\
N^1_2&=E[N_2]=\frac{\log M_1}{R_1^*}+
\frac{\log M_2}{(C_2-\epsilon)}-\frac{R_2^*}{(C_2-\epsilon)}
\frac{\log M_1}{R_1^*}\text{  a.s.}
\end{align*}
this implies that
\begin{align*}
\frac{\log M_1}{E[N_1^1]}&=R_1^*\\
\frac{\log M_2}{E[N_2^1]}&=\frac{N_1^1}{N_2^1}R_2^*+
(1-\frac{N_1^1}{N_2^1})(C_2-\epsilon).
\end{align*}

Thus, by letting $N_1^1$ and $N_2^1$ be arbitrary large
with $\frac{N_1^1}{N_2^1}= r_2$,
this coding scheme achieves any rate pair within the outer region.
In the case where $E[N]=E[N_2]$, a symmetric construction shows that
the outer region of Theorem \ref{th:convmac} is achieved.
Let us denote by  ${\cal V}^2$ the $(M_1,M_2,N^2_1,N^2_2)$ variable length code
corresponding to this construction.

This shows that, in the particular cases where $E[N]=E[N_1]$ or
$E[N]=E[N_2]$, the best coding scheme is composed of two successive
block codes. Hence, in this example,
we see that the gain in terms of achievable rates essentially comes from
the possibility for the receiver to decode each message at
a different instant of time.

Concerning the general case with no specific restriction on $E[N]$,
for a fixed value of $E[N_1]$ and $E[N_2]$,
the best outer bound is obtained by minimizing $E[N]$.
Since $N_1\geq\frac{\log M_1}{C_1}$ and $N_2\geq\frac{\log M_2}{C_2}$
with high probability, we have $E[N]\geq\min(\frac{\log M_1}{C_1},
\frac{\log M_2}{C_2})$. 
In the remaining of this section, we restrict our analysis to  
coding schemes with $\frac{\log M_1}{C_1}=\frac{\log M_2}{C_2}$,
this impose a restriction on the ratio of the expected decoding times.
For such codes, using the lower bound on $E[N]$, we have 
$1\geq r_1\geq\frac{R_1}{C_1}$ 
and $1\geq r_2\geq\frac{R_2}{C_2}$,
thus we may rewrite the outer bound on ${\cal C}_{r_1,r_2}$
for these values of $r_1$ and $r_2$, as
\begin{align*}
R_1&\leq\frac{C_1}{2-\frac{I(X_1;Y|X_2,Q)}{C_1}}\\
R_2&\leq\frac{C_2}{2-\frac{I(X_2;Y|X_1,Q)}{C_2}}\\
sR_1+R_2&\leq \frac{R_2}{C_2} I(X_1,X_2;Y|Q)\\ 
&\quad + (1-\frac{R_1}{C_1})s C_1
+(1-\frac{R_2}{C_2})C_2,
\end{align*}
for some joint distribution $p(q)p(x_1|q)p(x_2|q)p(y|x_1,x_2)$,
and for $s=\frac{R_2}{R_1}\frac{C_1}{C_2}$. 
The last inequality can be worked out to bound
$R_2$ by a function of $R_1$, % and reciprocally.
%\begin{align*}
%R_2\leq\frac{C_2+sC_1-s2R_1}{2-\frac{I(X_1,X_2;Y|Q)}{C_2}}.
%\end{align*}
%Assuming furthermore that $s=\frac{R_2}{R_1}$, that is $\log M_1=\log M_2$ 
%and $C_1=C_2\triangleq C$, we get 
\begin{align*}
R_2\leq\frac{C_2}{2+2\frac{C_1}{C_2}-\frac{C_1^2}{C_2R_1}
-\frac{I(X_1,X_2;Y|Q)}{C_2}}.
\end{align*}
Thus, when $R_1$ satisfies its upper bound with equality,
$R_2$ must satisfy
\begin{align*}
%R_2\leq\frac{C}{2-\frac{I(X_1,X_2;Y|Q)-I(X_1;Y|X_2,Q)}{C}},
R_2\leq\frac{C_2}{2-\frac{I(X_2;Y|Q)}{C_2}},
\end{align*}
for some joint distribution $p(q)p(x_1|q)p(x_2|q)p(y|x_1,x_2)$.

\begin{figure}[t]
%\vspace{.5\baselineskip}
\centering
\includegraphics{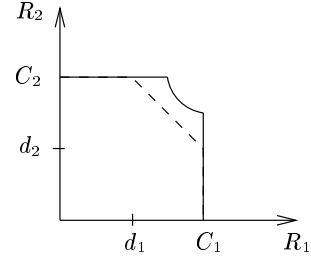}
\caption{Example of a region of achievable rates for 
a multiple-access channel.
The figure shows the region ${\cal C}_{r_1,r_2}$ for a multiple-access
channel with a pentagon-shape capacity region and for variable length
codes for which $\frac{\log M_1}{C_1}=\frac{\log M_2}{C_2}$.  
The dashed line delimits the achievable region using block codes.}
\label{fig:macregion}
\end{figure}

We specify now this outer region when the block code 
capacity region of the multiple-access channel forms a pentagon.
%Let $d_1=I(X_1;Y)$ and $d_2=I(X_2;Y)$
%and let $I(X_1;Y|X_2,Q)=d_1+p(C_1-d_1)$ 
%for some $p\in[0,1]$, be on the dominant face of ${\cal R}_{MAC}$, 
Let us denote by $(C_1,d_2)$ and $(d_1,C_2)$ the corner points of 
the dominant face of ${\cal R}_{MAC}$ (see Fig. \ref{fig:macregion}).
Then, let the joint distribution be such that the pair 
$(I(X_1;Y|X_2,Q),$
$I(X_2;Y|Q))$ is on the dominant face of ${\cal R}_{MAC}$,
we can describe any such pair by 
$I(X_1;Y|X_2,Q)=d_1+p(C_1-d_1)$ and 
$I(X_2;Y|Q)=d_2$
$+\bar p(C_2-d_2)$, for some $p\in[0,1]$.
Therefore, in this setting, the achievable rates satisfy
\begin{align}
\label{eq:mac_outer_region}
R_1&\leq\frac{C_1}{1+\bar p(1-\frac{d_1}{C_1})}\nonumber\\
R_2&\leq\frac{C_2}{1+p(1-\frac{d_2}{C_2})},
\end{align}
for some $p\in[0,1]$.
Observe that the region of all rate pair satisfying 
\eqref{eq:mac_outer_region} is not convex.

In order to achieve this bound, 
we consider variable length codes with non-deterministic encoders.%
\footnote{Note that, our setting can be extended to incorporate 
non-deterministic encoders and the outer bound on ${\cal C}_{r_1,r_2}$
still holds.}  
The idea is to use the codes ${\cal V}^1$ and ${\cal V}^2$
in alternation. 
To communicate a message pair $(w_1,w_2)\in (W_1,W_2)$, 
with probability $\lambda$, the transmitters use the codeword pair in 
${\cal V}^1$ corresponding to $(w_1,w_2)$, 
and with probability $1-\lambda\triangleq \bar \lambda$
they use the corresponding codeword pair in ${\cal V}^2$.
The codewords obtained by this procedure
form the codebook which is revealed to the receiver (and the transmitters). 
%Notice that the receiver knows which code is used by the transmitters.
This is a kind of ``time-sharing'' between the codes ${\cal V}^1$ 
and ${\cal V}^2$, except that
here the two codebooks have a different timeliness, and thus
we cannot construct a new codebook with the desired rates
by simply using one codebook a fraction of time and the 
other the remaining fraction of time.
The decoding times $N_1$ and $N_2$ of this coding scheme satisfy
\begin{align*}
E[N_1]=\lambda N_1^1+\bar \lambda N_1^2\\
E[N_2]=\lambda N_2^1+\bar \lambda N_2^2.
\end{align*}
%For $M_1$ and $M_2$ arbitrary large,
Now, for some $\epsilon>0$, set $(R_1^*,R_2^*)=(C_1-\epsilon,d_2)$ 
in the first block code of ${\cal V}^1$,
and $(R_1^*,R_2^*)=(d_1,C_2-\epsilon)$ in the first block code of ${\cal V}^2$.
For $E[N_1]$ and $E[N_2]$ arbitrary large,
this random coding scheme achieve the following rates
\begin{align*}
R_1&
%=\frac{\log M_1}{\frac{\log M_1}{C_1}+\bar \lambda(\frac{\log M_2}{C_2}
%-\frac{d_1}{C_1}\frac{\log M_2}{C_2})}
=\frac{(C_1-\epsilon)}{1+\bar \lambda \frac{\log M_2}{\log M_1}
\frac{(C_1-\epsilon)}{(C_2-\epsilon)}(1-\frac{d_1}{(C_1-\epsilon)})}\\
R_2&
%=\frac{\log M_2}{\frac{\log M_2}{C_2}+\lambda(\frac{\log M_1}{C_1}
%-\frac{d_2}{C_2}\frac{\log M_1}{C_1})}
=\frac{(C_2-\epsilon)}{1+\lambda \frac{\log M_1}{\log M_2}
\frac{(C_2-\epsilon)}{(C_1-\epsilon)}(1-\frac{d_2}{(C_2-\epsilon)})},
\end{align*}
for all $\lambda\in[0,1]$.

This can be related to the outer bound given by \eqref{eq:mac_outer_region}, 
in particular
for $\frac{\log M_1}{C_1}=\frac{\log M_2}{C_2}$, we have that
any rate pair $(R_1,R_2)$ such that
\begin{align*}
R_1&\leq\frac{(C_1-\epsilon)}{1+\bar \lambda(1-\frac{d_1}{(C_1-\epsilon)})}\\
R_2&\leq\frac{(C_2-\epsilon)}{1+\lambda(1-\frac{d_2}{(C_2-\epsilon)})},
\end{align*}
for some $\lambda\in[0,1]$, is achievable.
Thus, any rate pair within the outer region is achieved in this special case,
showing that ``time-sharing'' coding strategies are sufficient
for this setting.
The shape of such a region is represented in Fig. \ref{fig:macregion}.
Note that one can achieve higher rates with variable length coding
than the rates achievable with fixed length coding even when $E[N_1]=E[N_2]$.
This holds only because of the possibility
for the transmitters to send a part of their message
in non-overlapping periods of time.

Finally we remark that these coding strategies
need to fix the transmission rates (through the decoding times)
before generating
the codebook, thus each transmitter is aware of the rate
used by the other transmitters.
In the next section we show the existence of variable length
codes achieving the direct part of the block code capacity region of 
a multiple-access channel without requiring
a common agreement between the transmitters
(decentralized setting).

\section{Random Variable Length Codes}
%A $(M_1,M_2,N_1,N_2)$ random variable length code is
%a variable length code in which the transmitters
%employ a random codebook, 
In this section we analyze the rates achievable
when the transmitters employ a random codebook, 
that is the sequence of mappings 
$\{x_{1i}(W_1)\}_{i\geq 1}$ (resp. $\{x_{2i}(W_2)\}_{i\geq 1}$)
are $M_1$ (resp. $M_2$) random sequences of i.i.d. 
samples distributed according to a probability distribution $p(x_1)$
(resp. $p(x_2)$) defined over ${\mathcal X_1}$ (resp. ${\mathcal X_2}$).

\subsection{A joint decoding rule}
Let each transmitter start the transmission of 
a uniformly chosen codeword in the random codebook.
At time $n$, the decoder bases its decision on the sequence
of received values $y^n$. If we constrain the decoding times
$N_1$ and $N_2$ to be {\it equal},
the joint decoder that minimizes the probability of error 
will use a MAP (maximum a posteriori) rule and choose
the messages index $(w_1,w_2)$ maximizing the probability
that $(w_1,w_2)$ is transmitted knowing the received sequence.
Let%
\footnote{The fact that 
the decoder knows the realization of the random codewords
is implicit in the definition of $\tau(n)$.}
\begin{align*}
\tau(n)=\max_{w_1,w_2}\text{Pr}((w_1,w_2)\text{ is transmitted}|y^n),
\end{align*}
then the optimal joint decoder (the one that minimizes the
expected decoding time subject to a probability of error constraint) 
will make a decision at the time instant $n$ for which $\tau(n)$
exceeds a pre-determined threshold, and decode the messages $(w_1,w_2)$
achieving the maximum in the MAP rule.  

\begin{figure}[h]
\centering
\includegraphics{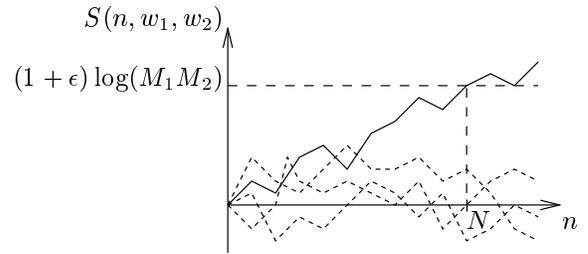}
\caption{Illustration of joint decoding with $M_1M_2=4$.
Each trace represents a random walk $\{S(n,w_1,w_2)\}_{n\geq 1}$
corresponding to a message pair $(w_1,w_2)$.
As soon as a random walk crosses the threshold given
by $(1+\epsilon)\log(M_1M_2)$, the decoder declares
the corresponding message pair.}\label{fig:rwalk}
\end{figure}

Since the optimal rule is difficult to analyze,
here we will make the hypothesis that
$p(y^n)=\Pi_{i=1}^n p(y_i)$,
and look at the following modified version
of the optimal decoding rule%
\footnote{
This assumption holds, since the channel is memoryless and 
we use a random codebook with i.i.d. samples,
but here the decoder knows the realization of
the random codewords and so he can compute all the conditional
probabilities such as
$p(y^n|x_1^n(1),\dots,x_1^n(M_1),x_2^n(1),\dots,x_2^n(M_2))$,
to obtain the true MAP rule.}
\begin{align*}
\tau(n)&=\max_{w_1,w_2}\frac{p(y^n|x^n_1(w_1),x^n_2(w_2))}{p(y^n)}\\
&=\max_{w_1,w_2}\Pi_{i=1}^n \frac{p(y_i|x_{1i}(w_1),x_{2i}(w_2))}{p(y_i)},
\end{align*}
taking the logarithm, we obtain
\begin{align*}
S_{joint}(n)&=\max_{w_1,w_2}
\sum_{i=1}^n\log\frac{p(y_i|x_{1i}(w_1),x_{2i}(w_2))}
{p(y_i)}.
\end{align*}
Let us denote the expression under the summation by 
\begin{align*}
Z_i(w_1,w_2)=\log\frac{p(y_i|x_{1i}(w_1),x_{2i}(w_2))}{p(y_i)}
\end{align*}
and the summation by $S(n,w_1,w_2)=\sum_{i=1}^n Z_i(w_1,w_2)$.
Note that for a fixed pair $(w_1,w_2)$, $\{Z_i(w_1,w_2)\}_{i\geq 1}$
is a sequence of i.i.d. random variables, and 
$\{S(n,w_1,w_2)\}_{n\geq 1}$ is a random walk.
Therefore, the joint decoder will declare the message pair $(w_1,w_2)$
corresponding to the first (among $M_1M_2$) random walk that 
crosses a given threshold (see Fig. \ref{fig:rwalk}).
Let us consider the following threshold $(1+\epsilon)\log (M_1M_2)$
with $\epsilon>0$, 
then $N$%
\footnote{Here we have $N_1=N_2=N$.}
is the stopping time defined by
\begin{align*}
N=\min \{n\geq 1:S_{joint}(n)\geq (1+\epsilon)\log(M_1M_2)\}.
\end{align*}
 
Assume, without lost of generality, that the message pair $(1,1)$ is transmitted,
and let us denote by $N_{1,1}$ the crossing time of the
random walk corresponding to the message pair $(1,1)$,
note that $N\leq N_{1,1}$.
Then, we have
\begin{align*}
E[Z_1(1,1)]=I(X_1,X_2;Y),
\end{align*}
using Wald's equality (see, e.g., \cite{G96}), we get
\begin{align*}
E[S(N_{1,1},1,1)]=I(X_1,X_2;Y)E[N_{1,1}].
\end{align*}
For $M_1M_2$ large we can ignore the overshoots and
$E[S(N_{1,1},1,1)]=(1+\epsilon)\log(M_1M_2)$.
Thus, we can conclude that
\begin{align*}
E[N]\leq E[N_{1,1}]\approx\frac{(1+\epsilon)\log(M_1M_2)}{I(X_1,X_2;Y)},
\end{align*}
which implies that 
\begin{align*}
R_1+R_2\geq \frac{I(X_1,X_2;Y)}{1+\epsilon}.
\end{align*}

The joint decoder makes an error when a random walk
corresponding to a different message pair crosses the threshold
before $\{S(n,1,1)\}$. 
The wrong messages come in three kinds:
\begin{enumerate}
\item $(w_1,w_2)$ such that $w_1\neq 1$ and $w_2\neq 1$,

\item $(w_1,w_2)$ such that $w_1=1$ and $w_2\neq 1$,

\item $(w_1,w_2)$ such that $w_1\neq 1$ and $w_2=1$.
\end{enumerate}
In each case we have
\begin{align*}
E[Z_1(w_1\neq1,w_2&\neq1)]\\
&=\sum_{x_1,x_2,y}p(x_1)p(x_2)p(y)
\log\frac{p(y|x_1,x_2)}{p(y)}\\
&=-D(p(y)||p(y|x_1,x_2))\\
&\leq 0,
\end{align*}
\begin{align*}
E[Z_1(w_1=1&,w_2\neq1)]\\
&=\sum_{x_1,x_2,y}p(x_1)p(x_2)p(y|x_1)
\log\frac{p(y|x_1,x_2)}{p(y)}\\
&\leq \sum_{x_1,x_2,y}p(x_1)p(x_2)\log\frac{p(y|x_1,x_2)}{p(y)}\\
&\leq \sum_{x_1,x_2,y}p(x_1)p(x_2)(\frac{p(y|x_1,x_2)}{p(y)}-1)\\
&=0,\\
%&=D(p(y|x_1)||p(y))-D(p(y|x_1)||p(y|x_1,x_2))\\
E[Z_1(w_1\neq1&,w_2=1)]\leq 0,
%D(p(y|x_2)||p(y))]-D(p(y|x_2)||p(y|x_1,x_2))\\
\end{align*}
where we use the fact that $\log x \leq (x-1)$,%
\footnote{Here ``$\log$'' denotes the logarithm to the base $e$.}
and the last inequality follows by symmetry.%
\footnote{Note that last two inequalities are equivalent to 
$D(p(y|x_1)||p(y))-D(p(y|x_1)||p(y|x_1,x_2))\leq 0$ and
$D(p(y|x_2)||p(y))-D(p(y|x_2)||p(y|x_1,x_2))\leq 0$,
with $p(y|x_1)=\sum_{x^{\prime}_2}p(x^{\prime}_2)p(y|x_1,x_2)$ and 
$p(y|x_2)=\sum_{x^{\prime}_1}p(x^{\prime}_1)p(y|x_1,x_2)$.}
Note that the expectations are taken with respect to the joint
probability $(X_1,X_2,Y)$ corresponding to the message pair $(w_1,w_2)$
considered. 

Thus $\{\{S(n,w_1,w_2)\}_{n\geq 1}:(w_1,w_2)\neq (1,1)\}$ are random walks
with negative drift. For those random walks one can show
(see, e.g., \cite{G96}) that the probability of ever crossing 
a threshold $T$ is upper bounded as follows
\begin{align*}
\text{Pr}(\text{crossing } T)\leq e^{-\lambda^*(w_1,w_2)T},
\end{align*}
where $\lambda^*(w_1,w_2)$ correspond to the unique positive root
of the log moment generating function of $Z_1(w_1,w_2)$, i.e.,
\begin{align*}
\log E[e^{\lambda^*(w_1,w_2)Z_1(w_1,w_2)}]=0.
\end{align*}
Therefore, we can upper bound the probability of error by the
probability that any random walk in 
$\{\{S(n,w_1,w_2)\}_{n\geq 1}:(w_1,w_2)\neq (1,1)\}$
crosses the threshold $T=(1+\epsilon)\log(M_1M_2)$:
\begin{align*}
P_e &\leq M_1 e^{-\lambda^*(w_1\neq 1,w_2=1)T}
+ M_2 e^{-\lambda^*(w_1= 1,w_2\neq1)T}\\
&\quad + M_1M_2 e^{-\lambda^*(w_1\neq 1,w_2\neq1)T}.
\end{align*}
Here we have
\begin{align*}
E[e^{Z_1(w_1\neq 1,w_2\neq 2)}]=
\sum_{x_1,x_2,y}p(x_1)p(x_2)p(y)\frac{p(y|x_1,x_2)}{p(y)}=1,
\end{align*}
which implies that $\lambda^*(w_1\neq 1,w_2\neq 1)=1$.
If, in addition
\begin{align*}
\lambda^*(w_1=1,w_2\neq1)&\geq \frac{I(X_2;Y|X_1)}{I(X_1,X_2;Y)}\\
\lambda^*(w_1\neq1,w_2=1)&\geq \frac{I(X_1;Y|X_2)}{I(X_1,X_2;Y)},
\end{align*}
we would have had
\begin{align*}
P_e &\leq e^{E[N](R_1-I(X_1;Y|X_2))}+e^{E[N](R_2-I(X_2;Y|X_1))}\\
&\quad +(M_1M_2)^{-\epsilon},
\end{align*}
and thus, by letting $\epsilon\rightarrow 0$ and $E[N]\rightarrow\infty$,
any rate pair in ${\cal R}_{MAC}$ with a fixed input
distribution $p(x_1)p(x_2)$ would have been achievable
using this joint decoding rule.
Unfortunately, in general, $\lambda^*(w_1=1,w_2\neq1)$ and 
$\lambda^*(w_1\neq1,w_2=1)$ do not satisfy the preceding inequalities.
However, we can improve this joint decoding scheme by combining it
with other schemes as explained
in the following subsection.

\subsection{A combined decoding rule}
\label{sec:comb}
We will combine the joint decoding rule
with the following decoding rules.
Suppose that the receiver knows which message the second transmitter
is sending, then the equivalent of the previous rule to decode the message coming
from the first transmitter is
\begin{align*}
S_{w_2}(n)&=\max_{w_1}\sum_{i=1}^n\log\frac{p(y_i|x_{1i}(w_1),x_{2i}(w_2))}
{p(y_i|x_{2i}(w_2))},
\end{align*}
where 
$p(y_i|x_{2i}(w_2))=\sum_{x^{\prime}_{1i}}p(x^{\prime}_{1i})p(y_i|x_{1i},x_{2i}(w_2))$.
Denote the expression under the summation by 
\begin{align*}
Z_i(w_1|w_2)=\log\frac{p(y_i|x_{1i}(w_1),x_{2i}(w_2))}{p(y_i|x_{2i}(w_2))}
\end{align*}
and the summation by $S(n,w_1|w_2)=\sum_{i=1}^n Z_i(w_1|w_2)$,
thus $\{S(n,w_1|w_2)\}_{n\geq 1}$ are $M_1$ random walks 
and the receiver will declare the message corresponding
to the first random walk crossing the pre-determined threshold.
Here, we let $N_{1,w_2}$ be the stopping time defined by
\begin{align*}
N_{1,w_2}=\min \{n\geq 1:S_{w_2}(n)\geq (1+\epsilon)\log M_1\}.
\end{align*}   
Assuming that the message pair $(1,1)$ is transmitted, 
we have
\begin{align*}
E[Z_1(1|1)]=I(X_1;Y|X_2),
\end{align*}
and
\begin{align*}
E[Z_1(w_1\neq1&|w_2=1)]\\
&=\sum_{x_1,x_2,y}p(x_1)p(x_2)p(y|x_2)
\log\frac{p(y|x_1,x_2)}{p(y|x_2)}\\
&=-D(p(y|x_2)||p(y|x_1,x_2))\\
&\leq 0.
\end{align*}
%and
%\begin{align*}
%E[Z_1(w_1\neq1|w_2\neq1)]&=\sum_{x_1,x_2,y}p(x_1)p(x_2)p(y)
%\log\frac{p(y|x_1,x_2)}{p(y|x_2)}\\
%&=-\sum_{x_1,x_2}p(x_1)p(x_2)[D(p_Y||p(\cdot|x_1,x_2))\\
%&\qquad\qquad-D(p_Y||p(\cdot|x_2))]\\
%&\leq 0.
%\end{align*}
As before we can upper bound the probability of error
knowing that the message $w_2=1$ is transmitted by 
the probability that a random walk corresponding to a different
message $w_1$ crosses the threshold $T_{w_2}=(1+\epsilon)\log M_1$,
which gives
\begin{align*}
P_{e,w_2}\leq M_1 e^{-\lambda^*(w_1\neq1|w_2=1)T_{w_2}},
\end{align*}
where $\lambda^*(w_1\neq1|w_2=1)$ is the unique positive root of 
the log moment generating function of $Z_1(w_1\neq1|w_2=1)$, 
which turns out to be equal to 1.
This allows us to conclude that 
\begin{align*}
E[N_{1,w_2}]\leq\frac{(1+\epsilon)\log M_1}{I(X_1;Y|X_2)},
\end{align*}
with
\begin{align*}
P_{e,w_2} &\leq M_1^{-\epsilon}.
\end{align*}
The same results hold if the receiver knows $w_1$ and wants to
decode $w_2$ (with an interchange 
of the indexes 1 and 2 on the above equations).

Now, let us remove the assumption that one of the transmitted message is
known by the receiver and combine these decoding schemes as follows. 
Consider a receiver which runs the three preceding 
decoding rules in parallel and 
declares the first message pair $(w_1,w_2)$ for which the corresponding
random walks have cross the threshold in each decoding scheme.
Such a decoder will run all the random walks $\{S(n,w_1,w_2)\}_{n\geq 1}$,
$\{S(n,w_1|w_2)\}_{n\geq 1}$ and $\{S(n,w_2|w_1)\}_{n\geq 1}$,
and stop when the random walks corresponding to one message pair
have hit the pre-determined threshold in each scheme,
that is the decoding time of this combined scheme is given by
\begin{align*}
N_{comb}=\min\{n\geq1:\exists (w_1,w_2) &\text{ and } n_1,n_2,n_3\leq n 
\text{ such that }\\
S(n_1,w_1,w_2)&\geq(1+\epsilon)\log (M_1M_2)\\
S(n_2,w_1|w_2)&\geq(1+\epsilon)\log M_1\\
S(n_3,w_2|w_1)&\geq(1+\epsilon)\log M_2
\}.
\end{align*}
%\begin{align*}
%N^{*}_{1,w_2}&=\min \{n\geq 1:\max_{w_2}S_{w_2}(n)\geq (1+\epsilon)\log M_1\}\\
%N^{*}_{2,w_1}&=\min \{n\geq 1:\max_{w_1}S_{w_1}(n)\geq (1+\epsilon)\log M_2\}.
%\end{align*}
This combined decoder will make an error when the random
walks corresponding to a wrong message pair will cross
the given threshold before the correct one in each scheme.
In regards of what has been said before, 
the probability of error of this combined decoder
can be bounded as follows
\begin{align*}
P_e &\leq M_1M_2e^{-\lambda^*(w_1\neq1,w_2\neq1)T}
+M_1e^{-\lambda^*(w_1\neq1|w_2=1)T_{w_2}}\\
&\quad + M_2e^{-\lambda^*(w_2\neq1|w_1=1)T_{w_1}},
\end{align*}
assuming that the message pair $(1,1)$ is transmitted.
Thus, we obtain that
\begin{align*}
P_e\leq (M_1M_2)^{-\epsilon} + M_1^{-\epsilon} + M_2^{-\epsilon},
\end{align*}  
and the probability of error goes to zero, as $M_1$ and $M_2$ get large. 
If we denote by $N_{1,1}$, $N_{1,w_=1}$ and $N_{2,w_1=1}$ the
crossing times of the random walks corresponding to the message $(1,1)$
in each of the three preceding schemes,
we can see from the expression of $N_{comb}$ that
\begin{align}
\label{eq:max_time}
E[N_{comb}]\leq E[\max(N_{1,1},N_{1,w_2=1},N_{2,w_1=1})].
\end{align}
At this point, let us remark that since the random walks
$S(n_1,1,1)$, $S(n_2,1|1)$ and $S(n_3,1|1)$ concentrate
around their mean as $n$ becomes large, 
the respective decoding times also concentrate around their mean
as the thresholds get large, this is show in Appendix \ref{app:con}.

Using this we see that as the crossing thresholds get large, 
each of the three preceding
decoding times concentrates around their mean and thus
the expectation in \eqref{eq:max_time} becomes
approximately equal to the maximum of the three 
expected decoding times, 
hence for $M_1$ and $M_2$ sufficiently large, we have
\begin{align}
\label{eq:max_time2}
E[N_{comb}]\leq\max(E[N_{1,1}],E[N_{1,w_2=1}],E[N_{2,w_1=1}]).
\end{align}
Therefore, for $M_1$ and $M_2$ large and for $\epsilon\rightarrow 0$, 
this random code approaches one of
the following rate pair
(in the same order that for the max in \eqref{eq:max_time2}),
depending on which expected decoding times 
is greater:
\begin{align*}
{\big(}\frac{\log M_1}
{\log M_1 + \log M_2}&I(X_1,X_2;Y),\\
&\frac{\log M_2}{\log M_1 + \log M_2}I(X_1,X_2;Y){\big)},
\end{align*}
\begin{align*}
&{\big(}I(X_1;Y|X_2),\frac{\log M_2}{\log M_1}I(X_1;Y|X_2){\big)},\\
&{\big(}\frac{\log M_1}{\log M_2}I(X_2;Y|X_1),I(X_2;Y|X_1){\big)}.
\end{align*}
Note that, according to the values of the ratio $\frac{\log M_1}{\log M_2}$,
any rate pair in ${\cal R}_{MAC}$ with a fixed input distribution $p(x_1)p(x_2)$
is achieved.
For example, if $M_1=M_2$ this coding scheme achieves the rate pair 
${\big(}\frac{I(X_1,X_2;Y)}{2}-\delta,\frac{I(X_1,X_2;Y)}{2}-\delta{\big)}$, and
if $\frac{\log M_1}{\log M_2}=\frac{I(X_1;Y|X_2)}{I(X_2;Y)}$
the rate pair achieved is ${\big(}I(X_1;Y|X_2)-\delta,I(X_2;Y)-\delta{\big)}$,
for some $\delta>0$.
Hence we have shown the existence of a variable length code
achieving a certain rate pair in ${\cal R}_{MAC}$,
without a previous agreement between the transmitters.
%However, due to the constraint on the decoding times
%a joint decoder will never be able to achieve rates outside
%${\cal R}_{MAC}$.

\subsection{A suboptimal decoding scheme}
\label{sec:subop}
We conclude this section by presenting a suboptimal
scheme that uses only single-user decoders, which is nothing but 
the successive decoding scheme adapted to variable length codes. 
Consider a receiver that decodes each message separately,
treating the signal of the other transmitter as noise.
In view of the preceding decoding rules, 
to decode the message of the first transmitter, 
we consider the following rule 
\begin{align*}
S(n)&=\max_{w_1}\sum_{i=1}^n\log\frac{p(y_i|x_{1i}(w_1))}
{p(y_i)},
\end{align*}
where 
$p(y_i|x_{1i}(w_1))=\sum_{x^{'}_{2i}}p(x^{'}_{2i})p(y_i|x_{1i}(w_1),x^{'}_{2i})$.
Denote the expression under the summation by 
\begin{align*}
Z_i(w_1)=\log\frac{p(y_i|x_{1i}(w_1))}{p(y_i)}
\end{align*}
and the summation by $S(n,w_1)=\sum_{i=1}^n Z_i(w_1)$,
then $\{S(n,w_1)\}_{n\geq 1}$ are $M_1$ random walks 
and the receiver will declare the message corresponding
to the first random walk crossing the pre-determined threshold.
Hence, we let $N_1$ be the stopping time defined by
\begin{align*}
N_1=\min \{n\geq 1:S(n)\geq (1+\epsilon)\log M_1\}.
\end{align*}  
Assuming that the message pair $(1,1)$ is transmitted, 
we have
\begin{align*}
E[Z_1(1)]=I(X_1;Y),
\end{align*}
and
\begin{align*}
E[Z_1(w_1\neq1)]&=\sum_{x_1,y}p(x_1)p(y)
\log\frac{p(y|x_1)}{p(y)}\\
&=-D(p(y)||p(y|x_1))\\
&\leq 0.
\end{align*}
As before we can upper bound the probability of error by 
the probability that a random walk corresponding to a different
message crosses the threshold $T_1=(1+\epsilon)\log M_1$,
which gives
\begin{align*}
P_e &\leq M_1 e^{-\lambda^*(w_1\neq 1)T_1},
\end{align*} 
where $\lambda^*(w_1\neq 1)=1$ is the unique positive root of the
log moment generating function of $Z_1(w_1\neq1)$. 
This allows us to conclude that 
\begin{align*}
P_e &\leq M_1^{-\epsilon},
\end{align*}
and for $M_1$ large
\begin{align*}
E[N_1]\leq\frac{(1+\epsilon)\log M_1}{I(X_1;Y)}.
\end{align*}
The same analysis apply to the decoding of the message sent
by the second transmitter. Thus, any rate pair $(R_1,R_2)$
such that $R_1<I(X_1;Y)$ and $R_2<I(X_2;Y)$ is achievable
using this strategy. 

Now let us improve this decoding scheme by noting that
as soon as one of the two messages are decoded,
the receiver can remove (the effect of) the signal of
the corresponding transmitter from the received signal.%
\footnote{In case of an additive channel this is done by a subtraction
which adds no complexity.}
Assume, without lost of generality, that the message from the second transmitter
is decoded earlier, then for decoding the message of the first 
transmitter, the receiver can use the rule $S_{w_2}$ previously
analyzed.
A receiver using this improved decoding rule is
able to decode the message coming from the first transmitter
at time $N_{1,w_2}$ and achieve any $R_1<I(X_1;Y|X_2)$. 
However, this decoding time might ``virtually'' happen
before $N_2$, the decoding time of the message coming 
from the second transmitter.
Thus, the actual decoding time of the message sent by 
the first transmitter is given by $\max(N_{1,w_2},N_2)$, 
which implies that in order to approach the rate pair 
$(R_1,R_2)=(I(X_1;Y|X_2),I(X_2;Y))$, the ratio $\frac{\log M_1}{\log M_2}$
%(which is equivalent to the ratio $\frac{E[N_1]}{E[N_2]}$)
must be sufficiently large.
%The same remark stands in the symmetric case.

\section{Concluding Remarks}
An explicit code approaching the transmission
rates of the random coding schemes presented here remains to be found.
%indeed the high complexity of our combined decoding rule
%suggests that practically good codes will be difficult to find. 
Nevertheless, for the suboptimal scheme presented in 
Section \ref{sec:subop}, and for certain multiple-access
channels, it might be interesting to 
consider coding schemes based on fountain codes. 
Notice that for the Gaussian multiple-access channel a practical
decoding scheme using rateless codes and successive 
decoding has been introduced in \cite{NESW06},
in the particular case where the cardinality of the
set of messages is the same for each transmitter
and when the decoding times are equal and deterministic.

Observe that our coding schemes can easily be adapted
to work when more than two users are simultaneously transmitting
and when the channel statistics are unknown to the transmitters,
as long as it is known to the receiver.
Furthermore, note that, in \cite{Shul03}, \cite{Tcham05} and \cite{TT06},
variable length codes are successfully used in
combination with different extension of the 
{\em maximum mutual information} (MMI) decoder,
to universally communicate over a class of unknown channels.
In the context of universal coding over a multiple-access channel,
the perfect mutual information decoder used in the
random coding schemes proposed here
may be replaced with the MMI decoder,
as done for the decoding strategies described in 
the above references.

Finally, we remark that the setup of this paper can be extended
to allow a noiseless and instantaneous
feedback from the receiver to the transmitters. This requires to
make each $x_{1i}$ and $x_{2i}$ dependent of the past received 
values $Y^{i-1}$. In this setting, we can prove the following
outer bound on ${\cal C}_{r_1,r_2}$,
if a rate pair $(R_1,R_2)$ is in ${\cal C}_{r_1,r_2}$, then
\begin{align*}
R_1&\leq r_1 I(X_1;Y|X_2) + (1-r_1)C_1\\
R_2&\leq r_2 I(X_2;Y|X_1) + (1-r_2)C_2\\
s R_1 + R_2 &\leq r_2 I(X_1,X_2;Y) + s(1-r_1)C_1 + (1-r_2)C_2, 
\end{align*} 
for some joint distribution $p(x_1,x_2)p(y|x_1,x_2)$.
This outer bound can easily be derived using the ideas
developed in the proof of Theorem \ref{th:convmac} 
(without the introduction of the time-sharing random variable $Q$).
This provides an extension of the outer bound on the capacity region 
of a multiple-access with feedback described in \cite{O84},
to the case where the receiver can decode the messages 
at different instants of time.

\appendices
\section{Proof of Lemma \ref{maclemma3}}
\label{app:lem}
Let $\xi_i = \text{1}\{N < i \leq N_1\}$,%
\footnote{As before, we define $Y_i\xi_i$
as being equal to $Y_i$ if $N<i\leq N_1$ and equal 
to $\aleph$ otherwise, where $\aleph$ denotes a
symbol distinct from any of the letters in $({\mathcal X_1},{\mathcal X_2},
{\mathcal Y})$.}
and consider
\begin{align*}
&I(W_1;Y_{N+1}^{N_1}|Y^{N},W_2)\\
%&= I(W_1;Y_1\xi_1,\cdots,Y_n\xi_n,\cdots|N,Y^N,W_2)\\
&= I(W_1;Y_1\xi_1,\xi_1,\cdots,Y_n\xi_n,\xi_n,\cdots|Y^N,W_2)
\end{align*}
\begin{align*}
&= I(W_1;\xi_1|Y^N,W_2)\\
&\quad + I(W_1;Y_1\xi_1|\xi_1,Y^N,W_2)+\cdots\\
&\quad + I(W_1;\xi_n|(Y\xi)^{n-1},\xi^{n-1},Y^N,W_2)\\ 
&\quad + I(W_1;Y_n\xi_n|(Y\xi)^{n-1},\xi^n,Y^N,W_2) + \cdots\\
&= \sum_{i=1}^{\infty} I(W_1;\xi_i|(Y\xi)^{i-1},\xi^{i-1},Y^N,W_2)\\
&\quad + \sum_{i=1}^{\infty} I(W_1;Y_i\xi_i|(Y\xi)^{i-1},\xi^i,Y^N,W_2),
\end{align*}
where we use the chain rule for mutual information
to obtain the second inequality. 
\\
\\
The first summation can be bounded as
\begin{align*}
\sum_{i=1}^{\infty} I(W_1;\xi_i|(Y\xi)^{i-1},\xi^{i-1},Y^N,W_2)
&\leq \sum_{i=1}^{\infty} H(\xi_i|\xi^{i-1})\\
&=H(\xi_1,\xi_2,\cdots)\\
&=H(N_1-N)\\
&\leq \log(eE[N_1-N]).
\end{align*}
where the last inequality is proved in \cite{C73}
and \cite[\textsection 1.3]{CK81}, as mentioned in the proof of Lemma 1.
\\
\\
For the second summation, we can write
\begin{align*}
I(W_1&;Y_i\xi_i|(Y\xi)^{i-1},\xi^i,Y^N,W_2)\\
&= H(Y_i\xi_i|(Y\xi)^{i-1},\xi^i,Y^N,W_2)\\
&\quad - H(Y_i\xi_i|(Y\xi)^{i-1},\xi^i,Y^N,W_2,W_1)\\
&\leq H(Y_i\xi_i|(X_{2i}\xi_i,\xi_i)\\
&\quad - H(Y_i\xi_i|X_{1i}\xi_i,X_{2i}\xi_i,(Y\xi)^{i-1},\xi^i,Y^N,W_2,W_1)\\
&\stackrel{(a)}{=} H(Y_i\xi_i|X_{2i}\xi_i,\xi_i)
- H(Y_i\xi_i|X_{1i}\xi_i,X_{2i}\xi_i,\xi_i)\\
&= \text{Pr}(\xi_i=1) {\big[} H(Y_i|X_{2i},\xi_i=1)\\
&\quad - H(Y_i|X_{1i},X_{2i},\xi_i=1) {\big]} \\
&= \text{Pr}(N < i \leq N_1) I(X_{1i};Y_i|X_{2i},\xi_i=1)\\
&\leq \text{Pr}(N < i \leq N_1) C_1,
\end{align*}
where in $(a)$ we use the fact that given $(X_{1i},X_{2i})$, 
$Y_i$ is independent of the past received values
and of $(W_1,W_2)$. The last inequality follows since 
$p(y_i|x_{1i},x_{2i},\xi_i=1)=p(y_i|x_{1i},x_{2i})$ and  
by the definition of $C_1$.
Thus, we get
\begin{align*}
&I(W_1;Y_{N+1}^{N_1}|Y^{N},W_2)\\
&\leq \log(eE[N_1-N]) + \sum_{i=1}^{\infty} \text{Pr}(N < i \leq N_1) C_1\\
&= \log(eE[N_1-N]) + E[N_1-N] C_1.
\end{align*}

\noindent
The second inequality follows in a symmetric way.

\section{}
\label{app:achiv}
Here, for channels with a zero-error capacity equal to zero,
we argue that the definition of achievability given by
Definition \ref{def:achiv} is equivalent to 
the following alternate definition of achievability.

\begin{definition}
A rate pair $(R_1,R_2)$ is said to be {\em achievable} if there exists
a sequence of $(M_1,M_2,N_1,N_2)$ variable length codes
with $E[N_1]$ and $E[N_2]$ increasing such that
$\liminf_{E[N_1]\rightarrow\infty,E[N_2]\rightarrow\infty}
P_e=0$.
\end{definition}

To see this, take the best variable length code
(the one that achieves the minimum $P_e$) 
with a finite $E[N_1]$ and/or $E[N_2]$ such that $\epsilon>P_e\geq\epsilon_1$,
for some $\epsilon>\epsilon_1>0$.
Note that $\epsilon_1$ could not be equal to zero otherwise this
would imply that the zero-error capacity of the channel is different than zero.
Hence, we can find an $\epsilon_2>0$ such that $\epsilon_1>\epsilon_2$.
Therefore, in order to achieve $P_e<\epsilon_2$, we need to increase $E[N_1]$
or $E[N_2]$. Repeating this argument, we see that $E[N_1]$ and $E[N_2]$
need to be arbitrarily large in order to achieve an arbitrary low
probability of error. 

\section{}
\label{app:con}
In this appendix, we show that for a random walk with a positive drift,
the time spend to hit a positive threshold 
concentrates around its mean.
Consider a random walk $S(n)=\sum_{i=1}^{n}Z_i$,
where $\{Z_i\}$ are i.i.d. random variables with $E[Z_1]>0$, 
and let $N$ be the first time at which $S(n)$ crosses a given threshold $T^*>0$.
By Wald's equality we know that for large $T^*$, 
$E[N]\approx\frac{T^*}{E[Z_1]}$, and here we want to show that
with high probability
$E[N](1-\epsilon^*)<N<E[N](1+\epsilon^*)$, for some $\epsilon^*>0$.
But, the following clearly holds
\begin{align*}
\text{Pr}(N\geq E[N](1+\epsilon^*))\leq\text{Pr}(S(E[N](1+\epsilon^*))\leq T^*),
\end{align*}
where the RHS corresponds 
to the probability that the random walk is 
under the threshold at time $E[N](1+\epsilon^*)$, which is
a large deviation event, since we have
\begin{align*}
&\text{Pr}(S(E[N](1+\epsilon^*))\leq T^*)\\
&=\text{Pr}{\big(}\frac{1}{E[N](1+\epsilon^*)}S(E[N](1+\epsilon^*))\leq 
\frac{E[Z_1]}{(1+\epsilon^*)}{\big)}\\
&\leq e^{-c(\epsilon^*)T^*},
\end{align*} 
where $c(\epsilon^*)$ is some constant depending on $\epsilon^*$.
The same conclusion can be obtained for the lower bound,
thus as $T^*$ gets large, $N$ concentrates around
its mean.

\section*{Acknowledgment}
%% optional entry into table of contents (if used)
%%\addcontentsline{toc}{section}{Acknowledgment}
The author wishes to thank Emre Telatar for insightful discussions
and helpful comments. 



% http://www.ctan.org/tex-archive/macros/latex/contrib/supported/IEEEtran/bibtex
%\bibliographystyle{IEEEtran.bst}

\begin{thebibliography}{1}

\bibitem{Burn76}
M. V. Burnashev,
``Data transmission over a discrete channel with
feedback: Random transmission time,''
{\sl Probl. Inf. Transm.}, vol. 12(4), pp. 250-265, 1976.

\bibitem{C98}
T. Cover,
``Comments on broadcast channels,''
{\sl IEEE Trans. Inf. Theory}, vol. 44, no. 6, pp. 2524-2530, Oct. 1998.

\bibitem{TC06}
T. Cover and J. Thomas,
{\sl Elements of Information Theory},
Wiley, New York, 2006.

\bibitem{C73}
I. Csisz\'ar,
``On the capacity of noisy channels with arbitrary signal costs,''
{\sl Problems of Control and Information Theory}, Vol. 2 (2-4), pp. 283-304, 1973.

\bibitem{CK81}
I. Csisz\'ar and J. K\"orner,
{\sl Information Theory: Coding Theorems for Discrete Memoryless Systems},
Academic Press, New York, 1981.

\bibitem{D78}
G. Dueck,
``Maximal error capacity regions are smaller than average error
capacity regions for multi-user channels,''
{\sl Prob. Contr. Inform. Theory}, vol 7, pp. 11-19, 1978.

%\bibitem{ES06}
%O.Etesami and A. Shokrollahi,
%``Raptor codes on binary memoryless symmetric channels,''
%{\sl IEEE Trans. Inf. Theory}, vol. 52, no. 5, pp. 2033-2051, May 2006.

\bibitem{G96}
R. G. Gallager,
{\sl Discrete Stochastic Processes},
Kluwer Academic Publishers, Boston, 1996.

\bibitem{L02}
M. Luby,
``LT codes,''
in {\sl proc. 43rd IEEE Symp. Foundations of Computer Science},
Vancouver, BC, Canada, Nov. 2002, pp. 271-280.

\bibitem{M07}
S. Musy,
``Variable length codes for degraded broadcast channels,''
in {\sl Proc. 2007 IEEE Int. Symp. Inf. Theory}, Nice, France, 
Jun. 2007, pp. 2576-2580.

\bibitem{NESW06}
U. Niesen, U. Erez, D. Shah and G. W. Wornell,
``Rateless codes for the Gaussian Multiple
Access Channel,''
in {\sl Proc. 2006 IEEE GLOBECOM}, San Francisco, CA, Nov. 2006.

\bibitem{O84}
L. H. Ozarow,
``The capacity of the white Gaussian multiple access channel 
with feedback,''
{\sl IEEE Trans. Inf. Theory}, vol. 30, no. 4, pp. 623-629, Jul. 1984.

\bibitem{S06}
A. Shokrollahi,
``Raptor codes,''
{\sl IEEE Trans. Inf. Theory}, vol. 52, no. 6, pp. 2551-2567, Jun. 2006. 

\bibitem{STV07}
S. Shamai, E. Telatar and S. Verd\'u,
``Fountain capacity,'' 
{\sl IEEE Trans. Inf. Theory}, vol. 53, no. 11, pp. 4372-4376, Nov. 2007.

\bibitem{Shul03}
N. Shulman, 
``Communication over an Unknown Channel via Common
Broadcasting,'' PhD thesis, Tel Aviv University, Jul. 2003.

\bibitem{SF00}
N. Shulman and M. Feder,
``Static broadcasting,''
in {\sl Proc. 2000 IEEE Int. Symp. Inf. Theory}, Sorrento, Italy,
Jun. 2000, p. 23.

\bibitem{Tcham05}
A. Tchamkerten,
``Feedback Communication over Unknown Channels,''
Ph.D. Thesis, Information Theory Laboratory, EPFL, Mar. 2005.

\bibitem{TT06}
A. Tchamkerten and E. Telatar,
``Variable length coding over an unknown channel,''
{\sl IEEE Trans. Inf. Theory}, vol. 52, no. 5, pp. 2126-2145, May 2006. 

\end{thebibliography}

\end{document}